\documentclass{ws-procs9x6}
\usepackage{graphicx}

\def\as{\ifmmode{{\alpha}_{s}}\else{${\alpha}_{s}$}\fi}
\def\rf{\ifmmode{{R}_{\rm 4}}\else{${R}_{\rm 4}$}\fi}
\def\ee{\mbox{e}^+\mbox{e}^-}

\begin{document}

\title{Determination of the strong coupling constant at
  LEP\footnote{Published in proceedings of DIS06, Tsukuba, Japan,
    20-24 April 2006.}}

\author{T. Wengler}

\address{University of Manchester \\
School of Physics and Astronomy \\ 
Manchester, M13 9PL, U.K.\\ 
E-mail: Thorsten.Wengler@manchester.ac.uk}

\maketitle

\vspace{-6.0cm}
\hspace{7.5cm} MAN/HEP/2006/22
\vspace{5.5cm}

\abstracts{ Multi-hadronic events produced in ${\ee}$ collisions
  provide an excellent laboratory to study QCD, the theory of strong
  interactions, and in particular to determine the strong coupling
  parameter {\as} and demonstrate its predicted behavior as a function
  of the energy scale. Determinations of {\as} at LEP will be reviewed
  with emphasis on event shape variables and jet rates in 3-jet and
  4-jet events.  }

\section{{\as} from 3-jet observables}

The grouping of particles into a number of collimated jets is one of
the most striking features of hadronic final states produced in
${\ee}$ collisions, and is easily reconcilable with the model of
energetic and hence boosted partons undergoing parton branchings and
hadronisation processes as prescribed by QCD. To quantify this
structure two types of observables are commonly used: event shapes and
jet rates.

\begin{figure}[t]
\includegraphics[bb=0 40 520 560, width=0.49\textwidth]{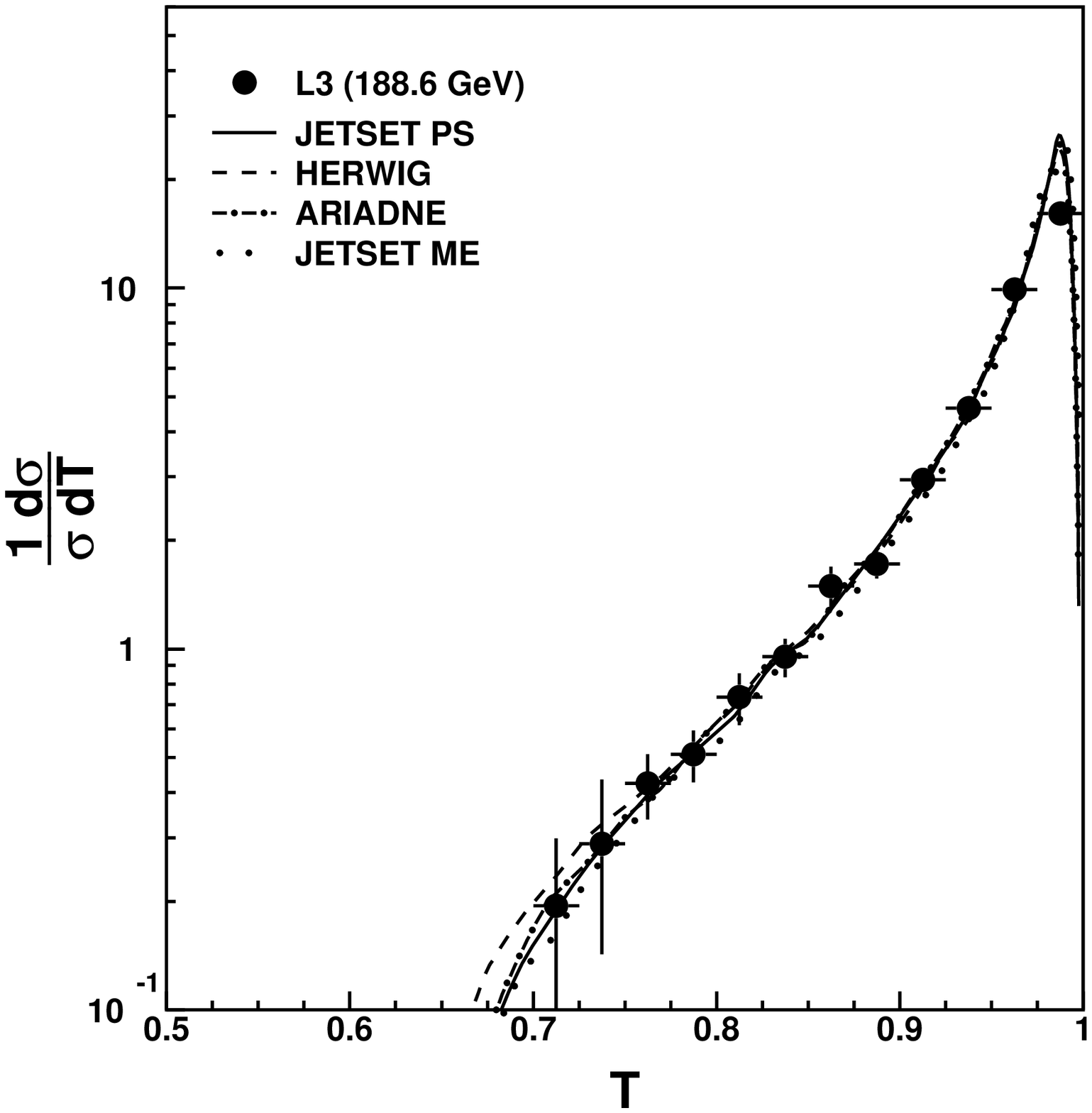}
\includegraphics[bb=0 0 565 565,width=0.49\textwidth]{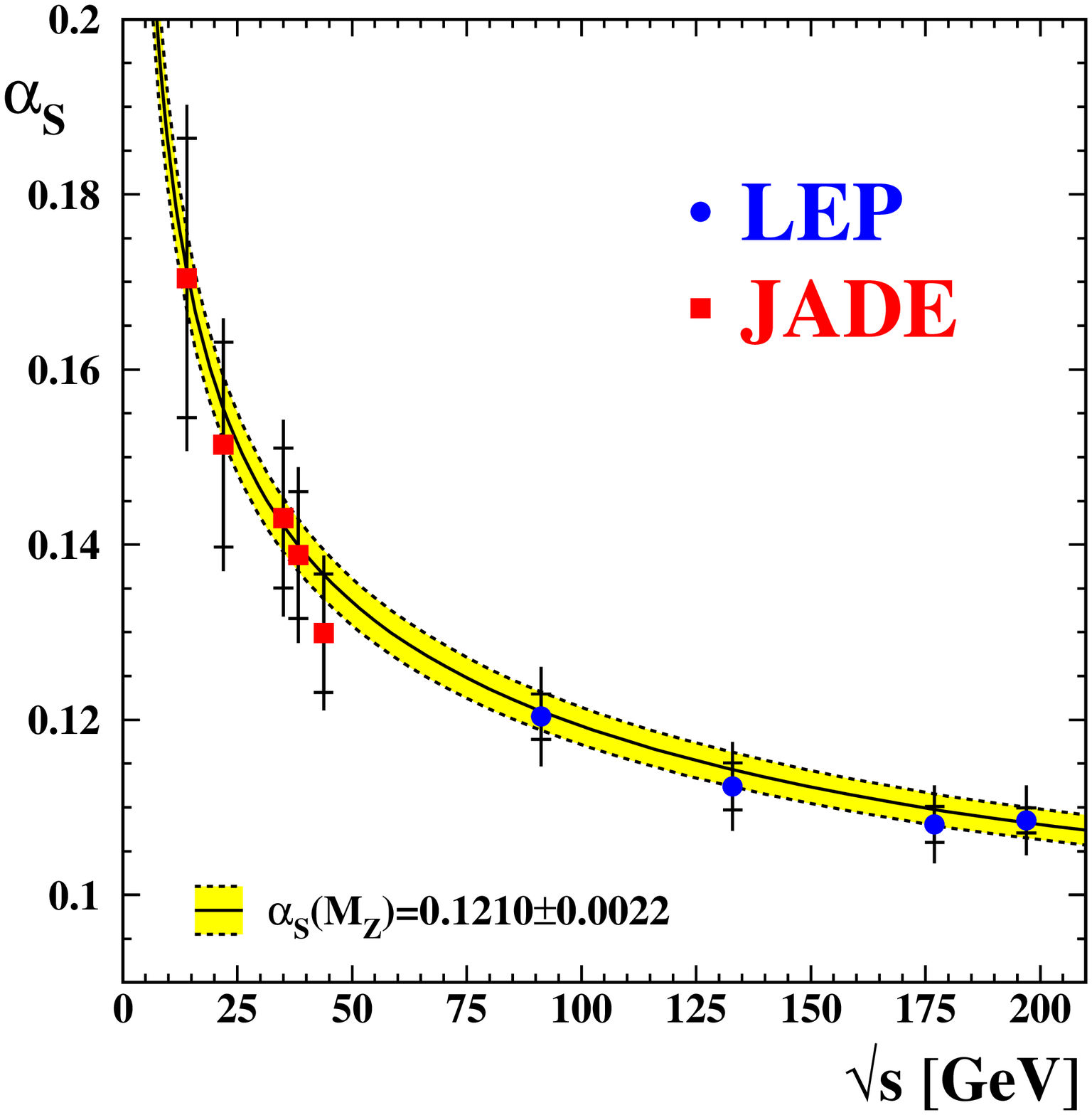}
\caption{The left picture shows an example of an event shape
  distribution - the Thrust, T, as measured by L3$^1$. The picture$^2$
  on the right summarises the combined values of {\as} based on the
  analysis of 3-jet observables by the LEP collaborations or using
  JADE data. \label{fig1}}
\end{figure}

To calculate jet rates clustering algorithms are used to group the
particles of the hadronic final state into a number of jets, based on
a resolution criterion which determines when the clustering should
stop. The rate of events with a given number of jets is directly
related to the coupling strengths involved. Event shapes on the other
hand are constructed by calculating a single number for each event
which classifies its topology. The left picture of Fig.~\ref{fig1}
shows the Thrust, $T$, as an example for an event shape
observable. The Thrust of an event is defined as the normalised sum of
absolute momentum components of all observed particles projected along
the axis that maximizes this sum. A well aligned, or ''pencil-like'',
2-jet event with few branchings will result in a value of $T$ close to
unity, while a more spherical event with many branchings will tend to
have smaller values of $T$.

The experimental procedure to determine the value of {\as} starts
with selecting multi-hadronic events, while rejecting events with
initial state radiation and $WW$ and $ZZ$ events. The measured
distributions are then corrected for detector effects, background and
efficiency, and theoretical predictions are fitted to determine
$\as$. The best available theoretical predictions involve
calculations in next-to-leading order (NLO) perturbative QCD matched
to next-to-leading-log (NLLA) resummed calculations.

The LEP experiments\cite{kluthqcd} have chosen six event shapes for
which NLO+NLLA calculations are available in addition to jet rates to
produce a combined value of $\as$: $(1-T)$, the heavy jet mass $M_H$,
the jet-broadening observables $B_T$ and $B_W$, the $C$-parameter, and
the value of the resolution parameter of the Durham jet algorithm that
marks the transition of a 2-jet event into a 3-jet event,
$y_{23}$. The value of {\as} determined at four LEP centre-of-mass
energies (CME) is shown in the right picture of Fig.~\ref{fig1}.

The yellow band in the right picture of Fig.~\ref{fig1} represents the
value of {\as} as determined from an NNLO analysis of inclusive
observables, like the properties of the $Z$ line shape or the ratio
of the longitudinal and total cross section. There is an excellent
agreement between these two methods, and also with values determined
from JADE data at lower energies. The overall combined value from
event shapes and 3-jet rates at LEP quoted at the $Z$ mass is
$\as(m_{\rm Z}) = 0.1201\pm 0.0053$, where the systematic error is
dominated by the theoretical uncertainty.

\section{{\as} from 4-jet observables}

\begin{figure}[t]
\includegraphics[bb=0 0 520 520, width=0.47\textwidth]{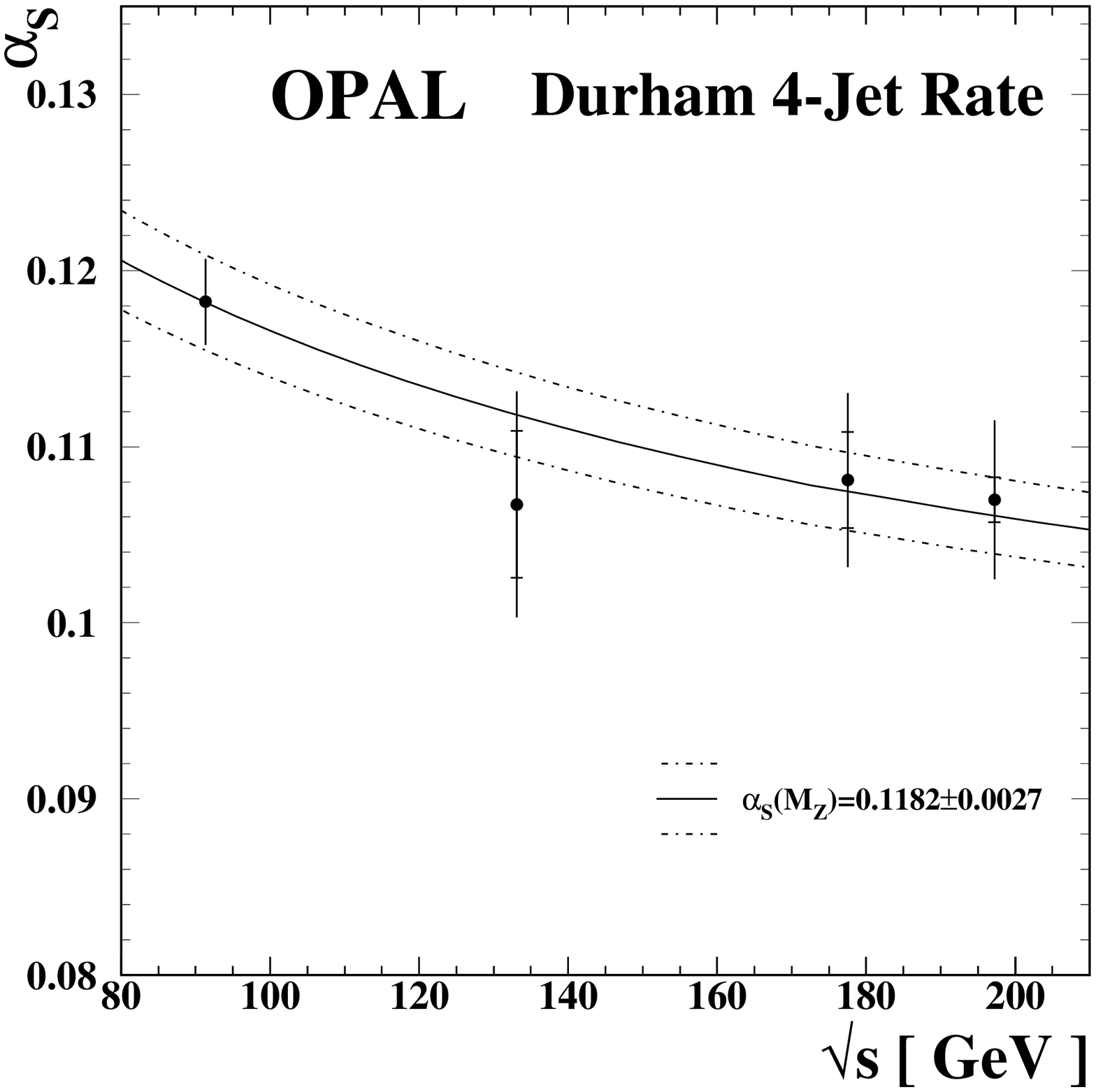}
\includegraphics[bb=0 18 570 552,width=0.49\textwidth]{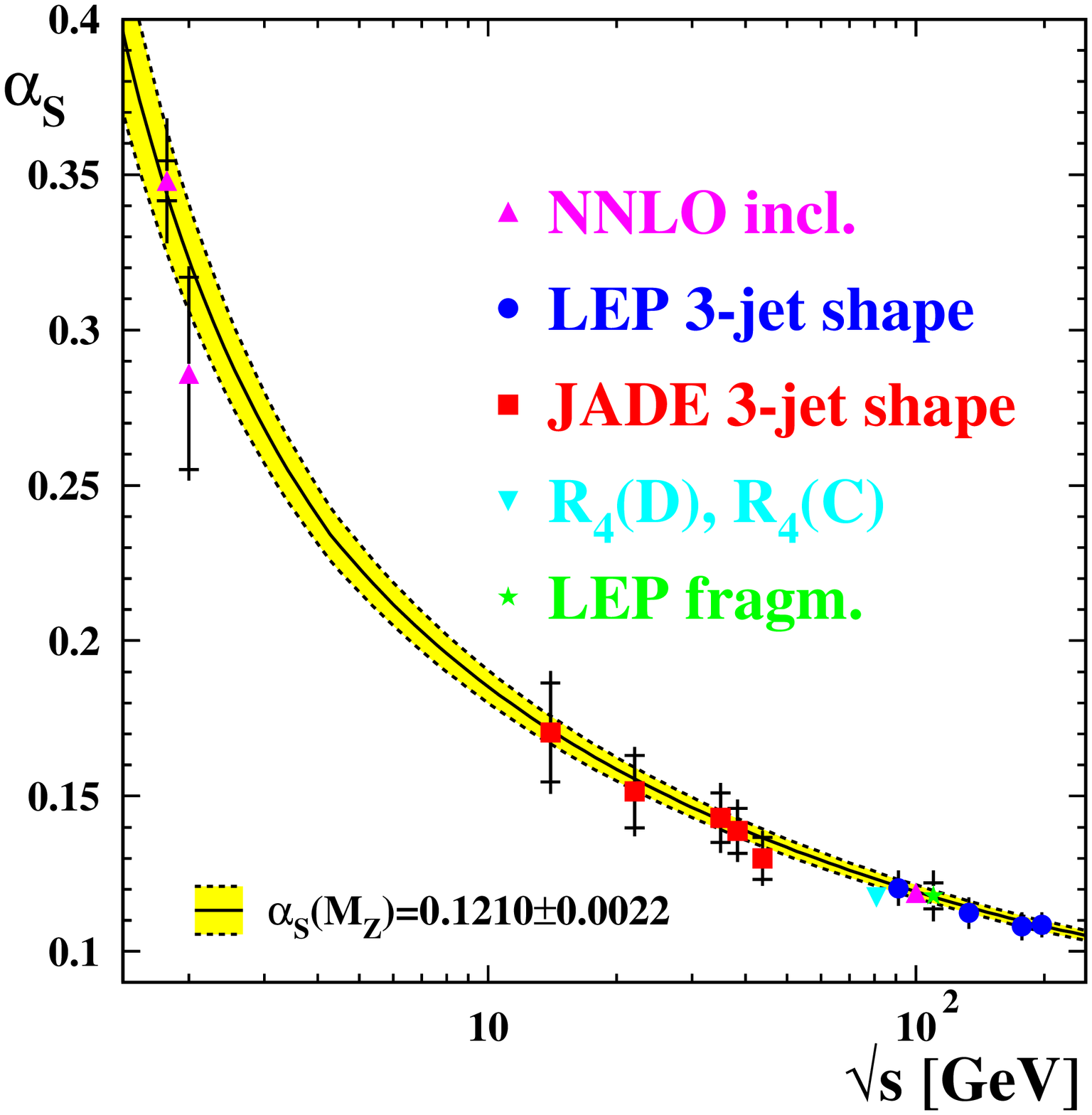}
\caption{The picture on the left shows the values of {\as} obtained by
  OPAL$^5$ from fits to the 4-jet rate in four intervals of the
  centre-of-mass energy ($\sqrt{s}$) at LEP2 . The picture$^2$ on the
  right is a summary of {\as} values determined at different values of
  $\sqrt{s}$. The yellow band represents the value of {\as} determined
  from inclusive variables at NNLO.\label{fig2}}
\end{figure}

Recently measurements have emerged from three of the four LEP
experiments using the 4-jet rate to determine {\as}. The 4-jet rate is
a promising observable, as its sensitivity to the value of {\as} is
double that of a 3-jet observable. On the other hand this means an
additional order of {\as} is needed in theoretical calculations to
reach NLO for this process, that is $O({\as}^3)$. Such calculations
are now available and have been matched with existing NLLA
calculations to produce the theoretical predictions needed to perform
the fits to the data.

ALEPH\cite{aleph4j} has fitted $O({\as}^3)$+NLLA calculations
corrected for hadronisation and detector effects to data at the $Z$
peak, yielding a value of $\as(m_{\rm Z}) = 0.1170\pm 0.0022$. The
uncertainty is dominated by theory. DELPHI\cite{delphi4j} has fitted
an $O({\as}^3)$ calculation corrected for hadronisation to data at the
$Z$ peak corrected for detector effects. Here no matching to an NLLA
calculation is attempted, but the renormalisation scale, $x_{\mu}$, is
optimised experimentally to reduce the influence of the theoretical
uncertainty. DELPHI determines $\as(m_{\rm Z}) = 0.1175\pm 0.003$,
with the uncertainty dominated by the hadronisation model, and not the
variation of the renormalisation scale, commonly used to assess the
theoretical uncertainty. It should be mentioned that an increased
variation of $x_{\mu}$ here leads to a drastic increase of the
theoretical uncertainty. OPAL\cite{opal4j} fits $O({\as}^3)$+NLLA
calculations corrected for hadronisation effects to data corrected for
detector effects from 91~GeV to 209~GeV CME. Values
of {\as} are presented at four CME points in the left
picture of Fig.~\ref{fig2}. Also shown is the central value and
uncertainty at the $Z$ mass resulting from a combination of the four
CME points: $\as(m_{\rm Z}) = 0.1182\pm 0.0026$, where the
uncertainty is dominated by theory.

A combination of all three results based on 4-jet rates as been
undertaken\cite{kluthqcd} and yields a value of $\as(m_{\rm Z}) =
0.1175\pm 0.0029$.

\section{Summary}

The value of the strong coupling {\as} has been determined at LEP
based on theoretically and experimentally well behaved observables of
event shapes and jet rates. {\as} determinations from 3-jet
observables yield reliable and precise results based on NLO+NLLA
calculations. The uncertainty is usually dominated by theory, and it
is hoped that theoretical developments will allow a reduction of the
uncertainty from now 5\% to 2\% in the near future. The first
determinations of {\as} from 4-jet rates based on NLO+NLLA
calculations are available and reach a precision comparable to the
most precise determinations today. Due to the small number of
measurements available so far the cross-checking of results is however
not yet as rigorous as achieved for the 3-jet observables. A summary
of {\as} determinations is shown in the right picture of
Fig.~\ref{fig2}.

A very consistent picture has developed across the various methods of
determining {\as} in $\ee$ collisions. The wide spread of measurements
in CME and the small uncertainties achieved for the
individual values allow a clear demonstration of the asymptotic
freedom of QCD.

\end{document}